\documentclass{aa}
\usepackage{graphicx}
\usepackage{amssymb}

\def\Msolar{\hbox{$M_\odot$}}

\begin{document}
\title{On the age and mass function of the globular cluster M\,4: a
different interpretation of recent deep HST observations\thanks{Based
on observations with the NASA/ESA {\it Hubble Space Telescope},
obtained at the Space Telescope Science Institute, which is operated by
AURA for NASA under contract NAS5-26555}}
 
\author{G. De Marchi \inst{1} \and F. Paresce \inst{2} \and O.
Straniero \inst{3} \and P.G. Prada Moroni \inst{4} \, \inst{5} \, \inst{6}}

\offprints{G. De Marchi}

\institute{European Space Agency, Space Telescope Operations Division,
           3700 San Martin Drive, Baltimore MD 21218, USA --
           \email{gdemarchi@rssd.esa.int}
\and
	   European Southern Observatory, Karl--Schwarzschild-Str. 2,
           85748 Garching, Germany -- \email{fparesce@eso.org}
\and
	   Osservatorio Astronomico di Teramo, Via Maggini, 64100
	   Teramo, Italy -- \email{straniero@te.astro.it}
\and
	   Dipartimento di Fisica, Universit\`a di Pisa, Via Buonarroti
	    2, 56127 Pisa, Italy -- \email{prada@df.unipi.it}
\and
	   Istituto Nazionale di Fisica Nucleare, Sezione di Pisa, 
	   56010 Pisa, Italy	
\and
	   Osservatorio Astronomico di Teramo, Via Maggini,
	   64100 Teramo, Italy
          }

\date{Received 3.6.2003 ; accepted 9.10.2003 }

\titlerunning{On the age of the globular cluster M\,4}
\authorrunning{De Marchi et al.}

\abstract{
Very deep images of the Galactic globular cluster M\,4 (NGC\,6121)
through the F606W and F814W filters were taken in 2001 with the WFPC2
on board the HST. A first published analysis of this data set (Richer
et al. 2002) produced the result that the age of M\,4 is $12.7\pm 0.7$
Gyr (Hansen et al. 2002), thus setting a robust lower limit to the age
of the universe. In view of the great astronomical importance of
getting this number right, we have subjected the same data set to the
simplest possible photometric analysis that completely avoids uncertain
assumptions about the origin of the detected sources. This analysis
clearly reveals both a thin main sequence, from which can be deduced
the deepest statistically complete mass function yet determined for a
globular cluster, and a white dwarf (WD) sequence extending all the way
down to the $5\,\sigma$ detection limit at $I \simeq 27$. The WD sequence
is abruptly terminated at exactly this limit as expected by detection
statistics. Using our most recent theoretical WD models (Prada Moroni
\& Straniero 2002) to obtain the expected WD sequence for different
ages in the observed bandpasses, we find that the data so far obtained
do not reach the peak of the WD luminosity function, thence only
allowing one to set a lower limit to the age of M\,4 of $\sim 9$\,Gyr.
Thus, the problem of determining the absolute age of a globular cluster
and, therefore, the onset of GC formation with cosmologically
significant accuracy remains completely open. Only observations several
magnitudes deeper than the limit obtained so far would allow one to
approach this objective.

}

\maketitle
 
\section{Introduction}

Stars in the metal-poor globular clusters (GCs) of our Galaxy's halo
are currently the oldest objects known whose age can be determined by
present means, thereby setting a firm lower limit on the age of the
Universe and allowing independent confirmation by other means of the
age recently obtained with WMAP (Bennett et al. 2003). The best
measurements so far based on main sequence (MS) fitting yield a value
of the age of the oldest GC of $12.5$\,Gyr with a $95\,\%$ confidence
range of $2.5$\,Gyr (Krauss 2001; Gratton et al. 2003), with the
largest contribution to the measurement error coming from the distance
uncertainty. Taken at face value, this number compares favourably with
the expansion age of the Universe implied by WMAP (Bennett et al. 2003)
and by the most recent $H_{\rm O}$ and geometry measurements ($13 \pm
3$\,Gyr; Lahav 2001) as well as with the results of radioactive dating
of a very metal-poor star ($12.5 \pm 3.0$\,Gyr; Cayrel et al. 2001).

A comparison between the measured expansion age of the universe
and the GC limit will also shed light on the presently uncertain and
controversial issue of the time of formation of the oldest halo GCs
(Gnedin, Lahav \& Rees 2001).  Current estimates of the epoch of
formation of these objects range from a minimum of $10-100$\,Myr, if
they were formed at the epoch of recombination, to a maximum of 5\,Gyr,
if they were formed as the result of thermal instabilities in the
Galactic halo (Fall \& Rees 1985). This enormous range can be
significantly reduced with a precise measurement of a GC age which, in
turn, might help us to determine, for example, whether the Galactic
halo formed from the accretion of dwarf galaxies or from protogalactic
cloud collapse (Mould 1998). What is really needed is a GC age
measurement in the range $10-15$\,Gyr with a $2\,\sigma$ uncertainty of
$10\,\%$ or less, comparable to the current precision on $H_{\rm O}$.

Since the main obstacle by far to a more precise determination of GC
ages lies in the uncertainty on their distance, the situation is
unlikely to change significantly until well after the launch of GAIA
and SIM in the next decade. In this regard, WDs could play an important
r\^ole both as distance indicators and as cosmo-chronometers and allow
measurements more accurate than the MS turn-off method. As recognised
early on by Mestel (1952), the decrease of the WD brightness with time
is the result of a cooling process so that the luminosity of a WD
indicates its age.  Unlike the turn off age--luminosity relationship,
the cooling timescale is independent of the original chemical
composition of the progenitor star. Cooling is generally rather fast,
except during the crystallisation of the core that lasts for several
Gyr.  Thus, a pile up of WDs is expected in the CMD of an old stellar
system. In practice, the WD luminosity function (LF) should present a
peak corresponding to the portion of WDs close to the end of their
crystallisation phase, followed by a sharp cutoff.  The luminosity of
this peak is a powerful age indicator that has been already used to
date the Galactic disc (Leggett et al.  1998) as well as some open
clusters (Von Hippel \& Gilmore 2000; Richer et al.  1998).  Owing to
the intrinsic faintness at which the WD LF peaks in very old stellar
systems, the extension of this method to most GCs has remained thus far
just a dream (see Fontaine, Brassard \& Bergeron 2001 for a detailed
discussion of the present state of the art). At least for the nearest
GCs, however, this goal is within the reach of the HST.  Indeed, using
deep observations of the WD cooling sequence of M\,4 obtained with the
WFPC2 camera, Hansen et al. (2002) have reported a determination of the
age of the cluster M\,4 to be 12.7\,Gyr to within $\pm 0.7$\,Gyr at the
$2\,\sigma$ level, thus attaining that accuracy needed to set
meaningful constraints on the age of the universe, as well as
concordance with it.

Obviously, reliable ages can be obtained only if an adequate
calibration of the age--luminosity relationship is available. This
problem has been recently reviewed by Prada Moroni \& Straniero (2002;
hereafter PMS02). They showed that, although in principle
cosmo-chronology based on WDs is a promising tool, in practice the
large discrepancies amongst the recently published theoretical cooling
sequences imply that a firm calibration of the age--luminosity
relationship is not yet available, especially for the range of ages
suitable for GCs.  The main reason for the quoted discrepancies is the
large uncertainty in the input physics needed to model the WD structure
and its evolution.  In particular, models depart progressively from one
another at low luminosities (see Figure\,1 in PMS02), where they are
very sensitive to the details of the physics of WD interiors and, thus,
provide ages that can vary by as much as 3\,Gyr at $\log
L/L_\odot=-5.5$. The key issue is that, since the cooling ages
predicted by different models begin to depart considerably from one
another for luminosities $\log L/L_\odot \lesssim -4.5$, any hope to
discriminate between competing theories by verifying the validity of
their predictions rests on our ability of securing a statistically
complete sample of WDs fainter than that luminosity.

We have, therefore, subjected the same data used by Hansen et al. 
(2002) to an independent scrutiny to verify at which level of 
significance they allow one to accept or reject a different set of WD 
cooling models, namely those of PMS02.

\section{The data}

The data used in this paper have been obtained with the WFPC2 on board
the HST as part of programme 8679 and are briefly described in Richer
et al. (2002). The target is a region located $\sim 5^\prime$\,E of
the nominal centre of M\,4 and has been imaged through the F606W filter
(98 images of duration 1300 s each) and in the F814W filter (148 images
each of duration 1300 s). Displacements of a fraction of a pixel
(dithering) have been applied between subsequent exposures in order to
improve the sampling of the point spread function (PSF) and to mitigate
the effects of cosmetic defects (Hook \& Fruchter 2000). We have
retrieved the complete dataset from the ESO/ST-ECF archive by making
use of the recalibration on-the-fly and automated association options
(Micol \& Durand 2002) to register and combine all images in the same
bandpass in a fully automated way. The accuracy of the registration and
combination procedures has been verified by comparing the properties of
the PSF in the individual images and in the combined ones. We find that
the full width at half maximum (FWHM) of stars in the frame increases
by about 15\,\%, as is typical of any shift-and-add operation conducted
on undersampled images. A known feature of this automated processing is
the incorrect subtraction of a constant sky from the final image, which
has, however, no effect on the accuracy of our measurements, since the
latter have been carried out via aperture photometry with the
background estimated locally in an annulus surrounding each object.

\begin{figure}
\resizebox{\hsize}{!}{\includegraphics{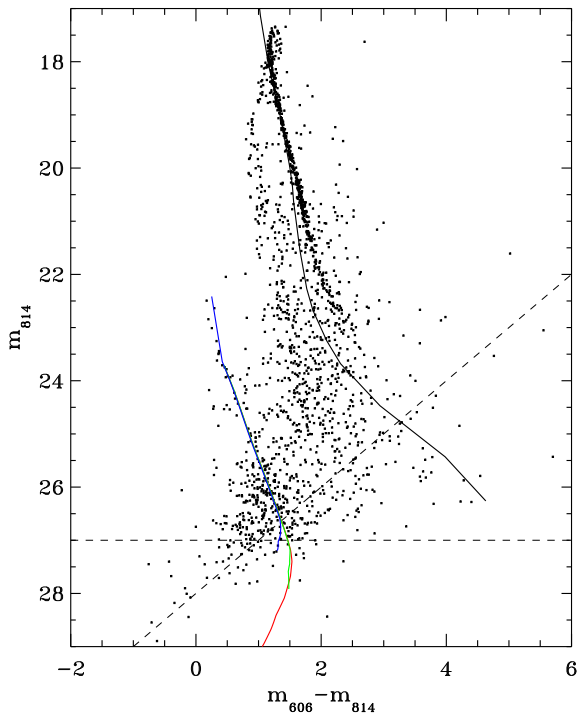}}
\caption{Colour--magnitude diagram of the three WF chips. The dashed
lines indicate the $5\,\sigma$ detection level, corresponding in
practice to 50\,\% completeness. The solid line next to the cluster MS
corresponds to the isochrone for an age of 10\,Gyr and metal content
$[M/H]=-1$ (Baraffe et al. 1997). The coloured solid lines drawn
through the WD region correspond to isochrones for ages of 8 (blue), 10
(green) and 12\,Gyr (red)}
\label{figure1}
\end{figure}

The equivalent exposure time of the final combined images (127,400 s in
F606W and 192,400 s in F814W) is such that stars with colour
$m_{606}-m_{814}\simeq 1$ have similar signal-to-noise ratio (SNR) in
both bands. To make the detection of both blue and red stars more
robust, we have registered the final images with one another and run
the IRAF {\it apphot.daofind} routine on the averaged frame. It is our
practice to conservatively set the detection threshold at $5\,\sigma$
above the local average background level to ensure that most noise
spikes and PSF artefacts are automatically rejected. In this case,
however, since we had combined the final F606W and F814W frames, we
decided to lower the limit to $4\,\sigma$ so as to avoid that very blue
or very red stars would be missed in the average image which could have
otherwise been detected at the $5\,\sigma$ level in at least one of the
two individual frames. We then carefully examined by eye each
individual object detected by daofind and discarded saturated stars, a
number of features even above $4\,\sigma$ (PSF tendrils, noise spikes,
etc.) that daofind had interpreted as stars, as well as several
extended objects (with FWHM larger than twice that typical of point
sources). The number of well defined objects detected in this way
amounts to $\sim 1730$ in the three WF chips (we did not reduce the PC
frame for it contains very few stars and would add minimally to the
statistics).

Crowding not being severe, stellar fluxes were measured by using the
standard {\it digiphot.apphot} IRAF aperture photometry routine,
following the prescription of the ``core aperture photometry''
technique described in De~Marchi et al. (1993). In particular, we
adopted an aperture radius of 2 pixel and a background annulus
extending from 3 to 5 pixel in radius. Aperture corrections
were calculated for an infinite aperture and the instrumental
magnitudes calibrated in the HST magnitude system (VEGAMAG)
by adopting the zero points listed in the January 2002 edition of the
HST Data Handbook (Mobasher et al. 2002). The photometric uncertainty
ranges from $\lesssim 0.01$ mag at $m_{606} \simeq 19$, $m_{814} \simeq
18$ to $\sim 0.35$ mag at $m_{606} \simeq 28.5$ and $\sim 0.25$
mag at $m_{814} \simeq 27$. The photometric uncertainties in each band, 
combined in quadrature, provide the error error on the colour.

To assess in a statistical way the completeness of our photometry, we
have run artificial star tests by adding, in repeated trials, several
fake stars of known brightness and subjecting each artificial image to
the same star finding and aperture photometry routines used for the
science frames. Because of the conservative detection threshold adopted
($4\,\sigma$ above the local background), an artificial star is always
recovered unless it overlaps with a brighter feature, such as another
star, an extended object, a spider ghost, a noise spike or other
feature. The completeness is, therefore, set at any magnitude by the
fraction of area which is free of brighter objects. The recovery
fraction is always larger than $\sim 95\,\%$ above $m_{814}=24.5$ and
slowly decreases to $\sim 80\,\%$ at $m_{814}=25.7$ and to $\sim
50\,\%$ at $m_{814}=26.8$. The difference between the input and
recovery magnitudes is consistent with the photometric errors quoted
above.

\section{The main sequence colour--magnitude diagram and luminosity function}

\begin{figure*}
\resizebox{\hsize}{!}{\includegraphics{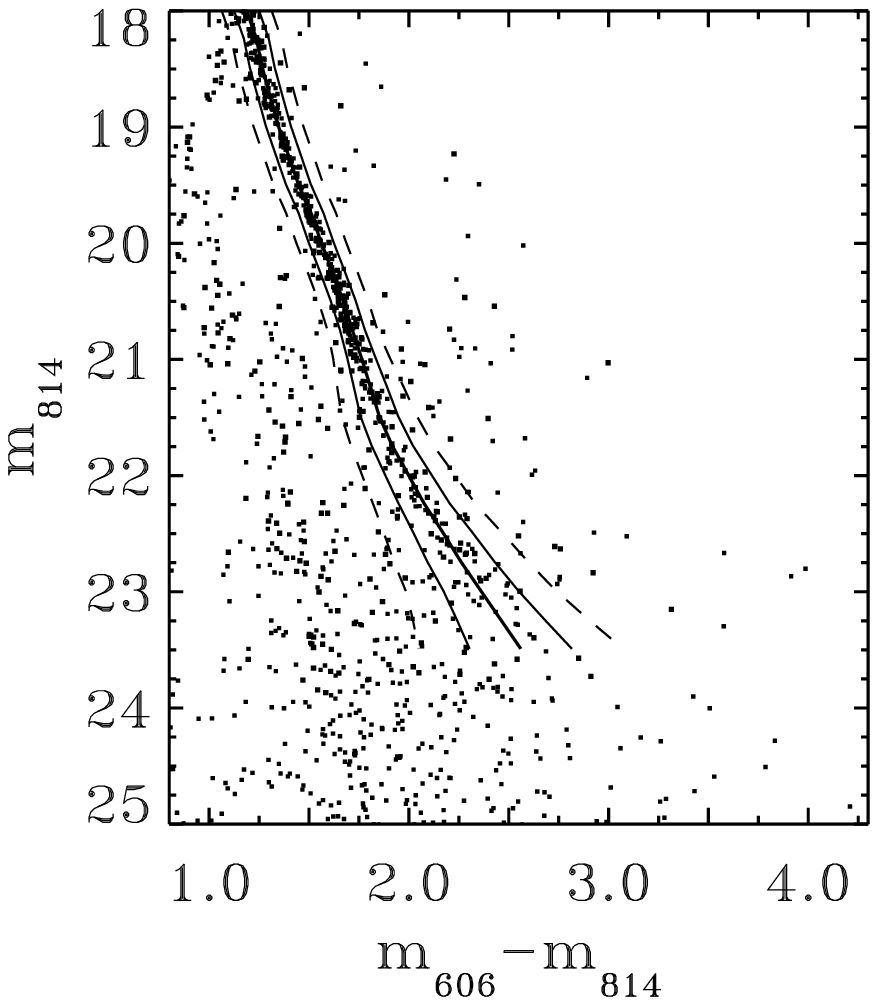}\hspace{0.5cm}
\includegraphics{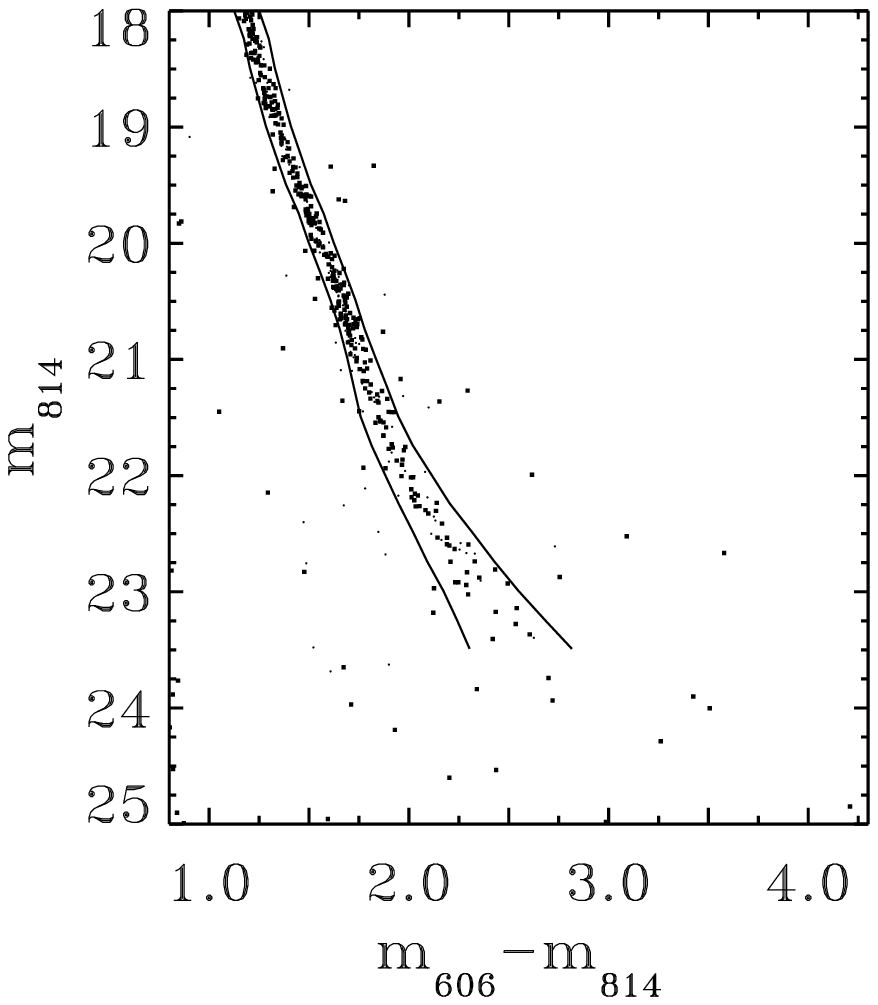}}
\caption{{\bf a.} (left panel): The thick solid line cutting through
the MS represents the ridge line obtained as explained in the text. To
measure the LF, we have counted the number of stars inside the thin
solid lines surrounding the MS ($\pm 4\,\sigma$ band) and subtracted
those found in the adjacent bands enclosed by the dashed lines (also
$4\,\sigma$ wide each).  {\bf b.} (right panel): Same as (a) but
showing only the stars with proper motion less than $0\farcs03$
(i.e. $0.3$ WF pixel) between the two epochs}
\label{figure2}
\end{figure*}

From the photometry obtained as explained in the previous section, we
derive the colour--magnitude diagram (CMD) shown in Figure\,1. The
dashed lines mark, for both bands, the magnitude limit at which the
photometry reaches the $50\,\%$ completeness level and happen to
correspond, in this particular case, to those objects whose central PSF
pixel rises above the local background level by 5 times the standard
deviation of the latter. This condition is, therefore, more stringent
than that set by the $4\,\sigma$ detection limit and suggests that we
exclude from the analysis that follows the objects below the dashed
lines, since their statistical significance would be anyhow too
uncertain. The cluster MS is narrow and well defined in
the range $ 17.5 \lesssim m_{814} \lesssim 23$. At the bright end, the
MS is truncated because of saturation (which, at least for the central
PSF pixel, extends up to $m_{814}\simeq 19.3$), whilst below
$m_{814} \simeq 23.5$ it becomes practically indistinguishable from
field stars (which, at that magnitude, span rather uniformly the colour
range $1 \lesssim m_{606} - m_{814} \lesssim 3$ to the left of the MS),
well before the onset of any significant appreciable photometric
incompleteness.

The solid line drawn next to the MS represents the theoretical
isochrone for an age of 10\,Gyr and metallicity of $[M/H]=-1$ as
obtained by Baraffe et al. (1997) in the specific bandpasses used here
for a distance modulus $(m-M)_V=12.83$ and colour excess $E(B-V)=0.36$
(Harris 1996), which in turn imply $(m-M)_I=12.25$.  As already noticed
by Bedin et al. (2001), who analysed shallower exposures of the same
area of M\,4 collected with the WFPC2 in the F555W and F814W filters,
the match between the observed and predicted MS is rather poor. They
attribute this discrepancy to the inadequate treatment of the
atmospheric opacity in the V band (F555W) compared to redder colours,
yet we find that the situation does not improve even when the redder
F606W band is used instead of the F555W one. By shortening the distance
modulus of about $\sim 0.5$ mag, one could better fit the
lower portion of the MS, but the isochrone would then fail to reproduce
the observations at the bright end, where models are known to be
accurate. Adopting a different set of theoretical models (Cassisi et
al. 2000; Montalban, D'Antona \& Mazzitelli 2000) does not
improve the fit. Certainly, a trustworthy isochrone drawn in
the plane of the observed colours would facilitate the selection of
true MS stars even at faint magnitudes, below $m_{814}\simeq 23$, where
the MS ridge line becomes increasingly uncertain. Sophisticated
techniques exist which use proper motion information from data
collected at different epochs to isolate bona-fide cluster objects from
field stars (see e.g. King et al. 1998 and Richer et al. 2002 for the
same dataset discussed here).  As we show in the following, even with this
technique, the sparseness of the MS and the lack of knowledge of its
colour hamper the characterisation of its lower end.

We use the colour and magnitude information contained in the CMD shown
in Figure\,1 to measure the LF for MS stars. We first determine the MS
ridge line by searching for the mode of the colour distribution as a
function of the magnitude. We use magnitude bins whose size increases
progressively from $0.2$\,mag at $m_{814}=19$ to $0.4$\,mag at
$m_{814}=23.5$ and a step along the magnitude axis equal to half the
size of the bin. The size of the bins has been selected in this way so
that the number of stars sampled is approximately constant, thus
minimising statistical fluctuations. An eye-ball fit through the ridge
line points found in this way is shown in Figure\,2\,a (thick solid
line). To count the number of stars along the MS we define a band,
around the ridge line, whose width is $\pm 4\,\sigma$ (thin solid
lines). For $m_{814}>21$, we take $\sigma$ to be the uncertainty on the
$m_{606}-m_{814}$ colour, whereas at brighter magnitudes we take
$\sigma = 0.013$, namely the colour uncertainty at $m_{814}=21$. To
account for field star contamination, the number of objects found
inside this band is reduced by the number of those detected within the
surrounding bands (dashed lines in Figure\,2\,a), each $4\,\sigma$ in
width.

\begin{figure}
\resizebox{\hsize}{!}{\includegraphics{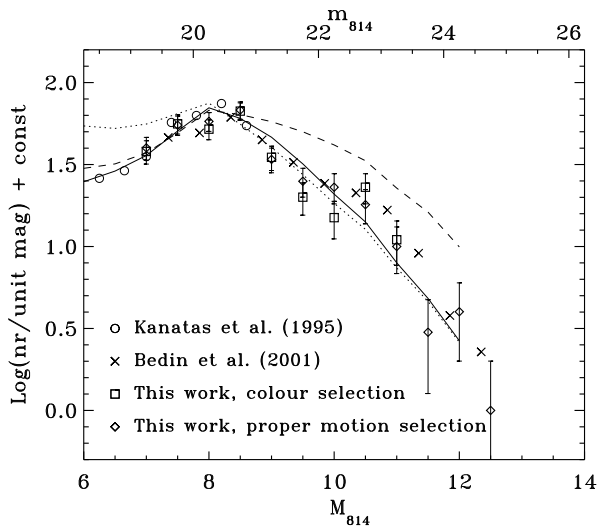}}
\caption{The data points show the LF of M\,4 as determined with
different methods and by different authors: squares, LF measured from
Figure\,2\,a; diamonds, LF measured from Figure\,2\,b; crosses, LF as
determined by Bedin et al. (2001) from shallower exposures of the same
field; plus signs, LF from Kanatas et al. (1995). The lines show
various MFs folded through the same M--L relationship (from Baraffe et
al. 1997): solid line, TPL with $m_p=0.35$\,\Msolar, $\alpha=2.1$ and
$\beta=2.7$; dashed line, power-law with $\alpha=0.75$; dotted line,
power-law with $\alpha=-0.3$.}
\label{figure3}
\end{figure}

The LF obtained in this way is plotted in Figure\,3 (squares) as a
function of the apparent (lower axis) and absolute (upper axis)
$m_{814}$ magnitude. No correction for incompleteness is applied, since
the latter is negligible, as stated in Section\,2. Error bars reflect
the Poisson statistics of the counting process, both on the MS and
contaminating stars. As previously noted, at magnitudes fainter than
$m_{814}\simeq 23.5$ the increasing photometric error as well as the
paucity of cluster stars in the CMD make it impossible to locate the
continuation of the MS amidst field stars.  For all practical purposes,
then, here the MS becomes statistically indistinguishable from field
stars and the LF goes to zero.

To overcome this statistical indetermination, we have retrieved from
the ESO/ECF archives a set of shallower images of the same cluster
region, obtained in 1995 with the WFPC2 in the F555W and F814W bands,
in an attempt to separate cluster and field stars through their proper
motion. These data, pertaining to HST programme 5461, are described in
Ibata et al. (1999). We have subjected these, first epoch, frames to
the same automated calibration and registration procedures used for the
deeper, second epoch, exposures described above, forming in this way
two combined frames with equivalent exposure time of 29,400 s in F555W
and 7,200 s in F814W, respectively. Due to the large proper motion of
M\,4, in the course of the six years between the two epochs cluster
objects should have moved, relative to background objects, by about
$0\farcs1$, or about a whole WF pixel. To measure this displacement, we
have followed the approach outlined in King et al. (1998) and have
determined the offset of each star in the reference frame defined by
some brighter ($19 < m_{814} < 21$) MS objects in its neighborhood. The
latter have initially been selected by virtue of their position in the
CMD but, through an iterative procedure, eventually only objects moving
less than $0.2$ pixel with respect to the average of their peers have
been retained as reference MS stars. The coordinates of each object are
those of its centroid as determined by the standard IRAF centering
procedure ({\it apphot.center}). They are very robust since, even near
the bottom of the MS, at $m_{814}\simeq 24$, the peak of the typical
star is detected at the $\sim 50\,\sigma$ and $\sim 20\,\sigma$ level
above the surrounding background, respectively in the first and second
epoch F814W combined frame. Furthermore, at both epochs the combined
images are well sampled thanks to the many dithered frames concurring
to form them.

Figure\,2\,b shows the CMD of those stars whose position has changed by
less than $0.3$\,pixel between the two epochs (the scatter on the
position of the bona-fide MS stars defined above being $\sim
0.1$\,pixel). Overplotted is the same MS band ($\pm 4\,\sigma$ wide)
shown in Figure\,2\,a, which was used to identify obvious outliers not
to be included in the counts to derive the LF. The latter is shown as
diamonds in Figure\,3, also not corrected for incompleteness, and
agrees remarkably well with that determined above (boxes) over the
common magnitude range. Below $m_{814}\simeq 23.5$, where the LF
derived with the MS-width method goes statistically to zero, there are
still a few data points in the CMD (Figure\,2\,b), which we have
assumed to be all MS objects if located more or less along the ideal
extension of the ridge line. Considering, however, that at about
$1.5$\,mag brighter there are some objects in the same colour range
which proper motion attributes to the cluster but which cannot be
considered MS stars, it is fair to wonder whether that assumption is
correct. For this reason, we consider the corresponding counts in the
LF as upper limits.

Also shown in Figure\,3 is the LF obtained by Bedin et al. (2001) by
combining the first epoch data discussed above with some shallower WFPC2
exposures (5,300 s long) in the F814W band alone, collected in the year
2000 (crosses). We do not plot their error bars so as not to clutter
the graph further, but the agreement is remarkable, thus independently
confirming the validity of proper motion studies, even when one of the
two epochs only provides information in one band.

Deriving the properties of the mass function (MF) underlying the LF
shown in Figure\,3 requires the use of a mass--luminosity (M--L)
relationship appropriate for the metallicity of the cluster. As
mentioned in Section\,2 (see Figure\,1), no isochrones today exist that
can reproduce the shape of the observed MS in the $V-I, I$ CMD. It is,
however, believed (Baraffe et al. 1998; Delfosse et al. 2000; Chabrier
2001) that the I band and near infrared colours are not affected by the
same shortcoming in the treatment of the atmospheric opacity mentioned
above because the latter is limited to visible colours. Thence, the
discrepancy shown in Figure\,1 should not prevent us from obtaining a
meaningful MF from the LF measured in the F814W band.

It is, unfortunately, customary to try and fit the observed LF by means
of a power-law MF. As Paresce \& De Marchi (2000a) have shown, however,
the stellar LF of most GCs measured to date with the HST reveals more
structure than that expected if the underlying MF were a simple
power-law. According to their analysis of 12 halo GC, the rapid drop of
the number counts seen in the LF at $m_{814} \gtrsim 8.5$ is not just
the result of the changing source of opacity in the stellar atmosphere
(D'Antona 1998) but an inflection in the MF is also needed at $\sim
0.3$\,\Msolar. Paresce \& De Marchi (2000a) propose that, over the
limited mass range covered by GC stars, a log-normal distribution
peaked at $\sim 0.3$\,\Msolar is more appropriate to describe the MF.
Alternatively, De Marchi, Paresce \& Portegies Zwart (2003; see also
Paresce \& De Marchi 2000b) propose a tapered power-law (TPL), namely a
Salpeter-like distribution which tapers off below a characteristic
mass. Analytically, the number of stars per unit mass can be expressed
as:

\begin{equation}
f \left(m \right)=\frac{d\,N}{d\,m}\propto m^{-\alpha} \left[ 1- 
e^{(-m/m_p)^\beta} \right]
\end{equation} 

\noindent
where $m_p$ is the peak mass, $\alpha$ the index of the power-law
portion for high masses and $\beta$ the tapering exponent which causes
the MF to flatten and drop below the peak mass $m_p$. For the 12 halo
GC studied by Paresce \& De Marchi (2000a), the average values of these
parameters are $\alpha=2.3$, $\beta=2.6$, $m_p=0.35$\,\Msolar. By
folding a function of this type through the M--L relationship of
Baraffe et al. (1997) for the metallicity $[M/H]=-1$, we are able to
reproduce rather well the LF over the entire range spanned by the
observations with $m_p=0.35$\,\Msolar, $\alpha=2.1 \pm 0.1$ and
$\beta=2.7 \pm 0.1$ (solid line in Figure\,3). It should be noted that
the shape of the LF in the magnitude range only accessible to the
proper motion technique does not set any constraints on the value of
these parameters, which are already defined by the brighter portion of
the LF.

It has been suggested by Richer et al. (2002) that the MF of M\,4 is
consistent with a power-law distribution of index $\alpha=0.75$ over
the entire range covered by these observations. A function of this
type, folded through the derivative of the Baraffe et al.'s (1997) M-L
relation, is shown as a dashed line in Figure\,3 and, regardless of the
arbitrary registration along the vertical axis, it does not provide a
good fit to the data. The dropping portion alone of the LF can be
represented with a power-law of index $\alpha=-0.3$ (dotted line in
Figure\,3). Because of the sign of its index, such a MF {\em increases}
with mass, thus implying that the number of stars per unit mass does
indeed decrease below $\sim 0.3$\,\Msolar. When this model MF is extended
to higher masses, however, it fails to reproduce the LF at $M_{814}
\lesssim 8$. To make this discrepancy more evident, we complement our
data at the bright end by showing, as circles, the LF of M\,4 as
measured by Kanatas et al. (1995) in the V band, after conversion to
the I band using the V, V-I CMD of Richer et al. (2002) as a
reference.

\section{The white dwarf cooling sequence and the age of the cluster}

Besides the MS, the other notable cluster feature in the CMD of
Figure\,1 is the WD cooling sequence, at magnitudes $m_{814} \gtrsim
23$ and colours $0 \lesssim m_{606}-m_{814} \lesssim 2$. The sequence
is sparse, yet narrow and well defined down to $m_{814}\simeq 25$,
where it broadens because of the increasing photometric uncertainty on
the colour, which rapidly grows from  $\sim 0.1$\,mag at $m_{814}=25$ to
$\sim 0.4$\,mag at $m_{814}=27$. The photometric completeness in this
region of the CMD decreases accordingly, as indicated above. The
sequence tapers off rapidly at $m_{814}\simeq 27$ where the
completeness reaches the 50\,\% level.

\begin{figure}
\resizebox{\hsize}{!}{\includegraphics{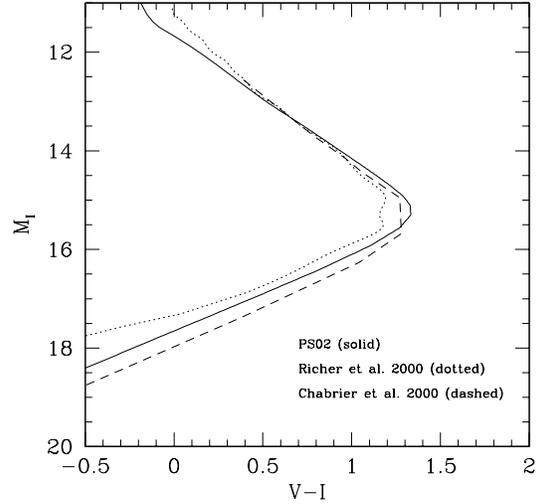}}
\caption{Evolutionary tracks of a $0.6$\,\Msolar WD as predicted by
the models of PMS02 (solid line), Richer et al. (2000; dotted line)
and Chabrier et al. (2000; dashed line). The filter names $V$ and $I$
here refer to the Johnson/Cousin photometric system}
\label{figure4}
\end{figure}

A comprehensive analysis of the observed WD cooling sequence requires
suitable theoretical models. We base our investigation on the DA WD
models recently computed by us (PMS02), which include the latest
improvements in the description of the physics of high density matter.
As discussed in that paper, theoretical uncertainties are still rather
large and they mainly affect the age--luminosity relationship for ages
in excess of  $6-7$\,Gyr. These uncertainties should be carefully
considered when dating an old WD sequence.

To translate the theoretical luminosity and temperature into magnitude
and colour for comparison with the HST observations, we used proper WD
model atmospheres computed by Bergeron, Leggett \& Ruiz (2001).  An
interesting prediction of the theory of the atmospheres of cool WD below
$T_{eff}\simeq 5000$\,K is the onset of collision-induced absorption,
due to H$_2$-H$_2$ and H$_2$-He collisions at such high densities
(Bergeron, Saumon \& Wesemael 1995; J\/orgensen et al. 2000). As
several authors have shown (Hansen 1998, 1999; Saumon \& Jacobson 1999;
Chabrier et al. 2000; Fontaine et al. 2001), the near infrared colours
of very cold WDs whose atmosphere contains Hydrogen should be
significantly affected by this source of absorption, with the peak of
their spectral energy distribution shifting to the blue.  A comparison
between our adopted evolutionary track of 0.6 M$_\odot$ and those
obtained by Richer et al. (2000) and Chabrier et al. (2000) is shown in
Figure\,4 for the Johnson/Cousin photometric system. All three
models  present an evident turn to the blue (``blue hook'') occurring
at nearly the same luminosity. The reddest colour is about $0.1$\,mag
bluer for the  sequence of Richer et al. (2000). After the ``blue
hook,'' the three sequences evolve at different luminosity, with that
of Richer et al. (2000) being the brightest and that of Chabrier et
al. (2000) the faintest. For example, at $V-I=0.5$ the relative
difference in the $I$ band is about half a magnitude. We note here
that, by adopting for M\,4 the same distance modulus used for MS
stars, namely $(m-M)_I=12.25$, the discrepancy between these three
theoretical cooling sequences is definitely smaller than the
photometric errors at the same magnitude level. In other words, the
present CMD does not permit a clear discrimination between the
different theoretical prescriptions and any conclusion should be
independent of the adopted set of WD models. In Section\,6 we elaborate
further on this issue.
 
The coloured lines over-plotted on the WD sequence in Figure\,1
represent the theoretical isochrones, as calculated from the WD models
of PMS02 for ages of 8, 10 and 12\,Gyr.  The initial--final mass
relationship and the age of the progenitors have been derived according
to the evolutionary models of low and intermediate mass stars with
Z=0.001 (Dominguez et al. 1999; Straniero et al. 1997), a metallicity
in agreement with the latest measurements\footnote{Carretta \& Gratton
(1997) find $[Fe/H]=-1.19 \pm 0.03$ and Carney (1996) gives
$[\alpha/Fe]=0.3 \pm 0.03$. This would imply $[M/H]=0.98$ or
$Z=0.0018$.} of heavy element abundances of giant stars in the field of
M\,4.

\begin{figure}
\resizebox{\hsize}{!}{\includegraphics{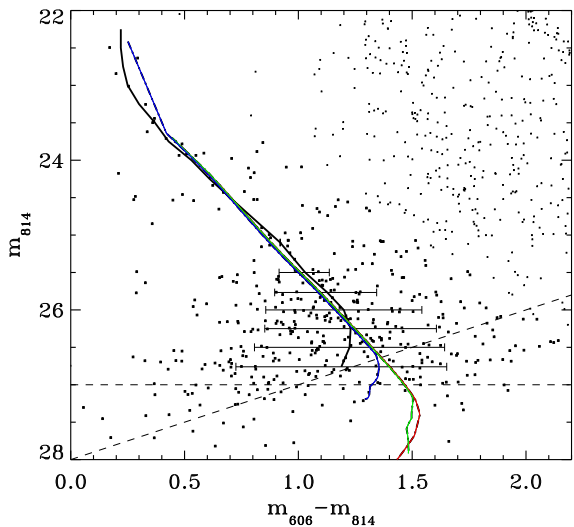}}
\caption{Enlargment of Figure\,1 showing the region occupied by the
WDs. The thick solid line cutting through the data shows the ridge line
of the WD cooling sequence estimated as explained in the text. As in
Figure\,1, the coloured lines mark the 50\,\% completeness limits.  The
WD isochrones for ages of 8 (blue), 10 (green) and 12\,Gyr (blue) are 
also shown}
\label{figure5}
\end{figure}

The three isochrones overlap one another over most of the magnitude
range. In a similar way as for the MS turn-off, the detection of the
``blue hook'' of the WD cooling sequence in the CMD could provide an
estimate of the cluster's age and would represent a much needed sanity
check as to the correctness of the models, thus allowing one to rely more
heavily on their predictions. Unfortunately, no such feature is visible
in Figure\,1 with any statistical significance. To prove this, we show
in Figure\,5 an enlargement of the WD region of the CMD of Figure\,1.
Although all data points within the indicated colour and magnitude 
limits are plotted, we only consider here those marked with a stronger
print and which comprise the WD region. The thick solid line cutting
through the data shows the ridge of the WD cooling sequence as
estimated by us.  Above $m_{814}=25.5$ this line is simply an eye-ball
fit to the points.  Below that threshold, however, any evidence of a
ridge suddenly vanishes, so there we plot the average colour as a
function of the magnitude, surrounded by its standard deviation (the
horizontal error bars). Below $m_{814}\simeq 26$, the average colour
appears to remain constant or even possibly bluer with increasing
magnitude. Given the conspicuous size of the error bars, however, this
inflection cannot be interpreted, with any statistical confidence, as
the presence of a ``blue hook.'' It could likely suggest a change in
the slope of the cooling sequence, but we should point out that
an even more likely origin for this effect can be traced in the onset
of considerable photometric incompleteness in the F606W band. As in
Figure\,1, the dashed lines in Figure\,5 mark the 50\,\% completeness
limits and it is clear that below $m_{814}\simeq 26$ an increasingly
larger fraction of redder objects must be missing, thus skewing the
distribution to the blue.

\begin{figure*}[t]
\centering
\resizebox{15cm}{!}{\includegraphics{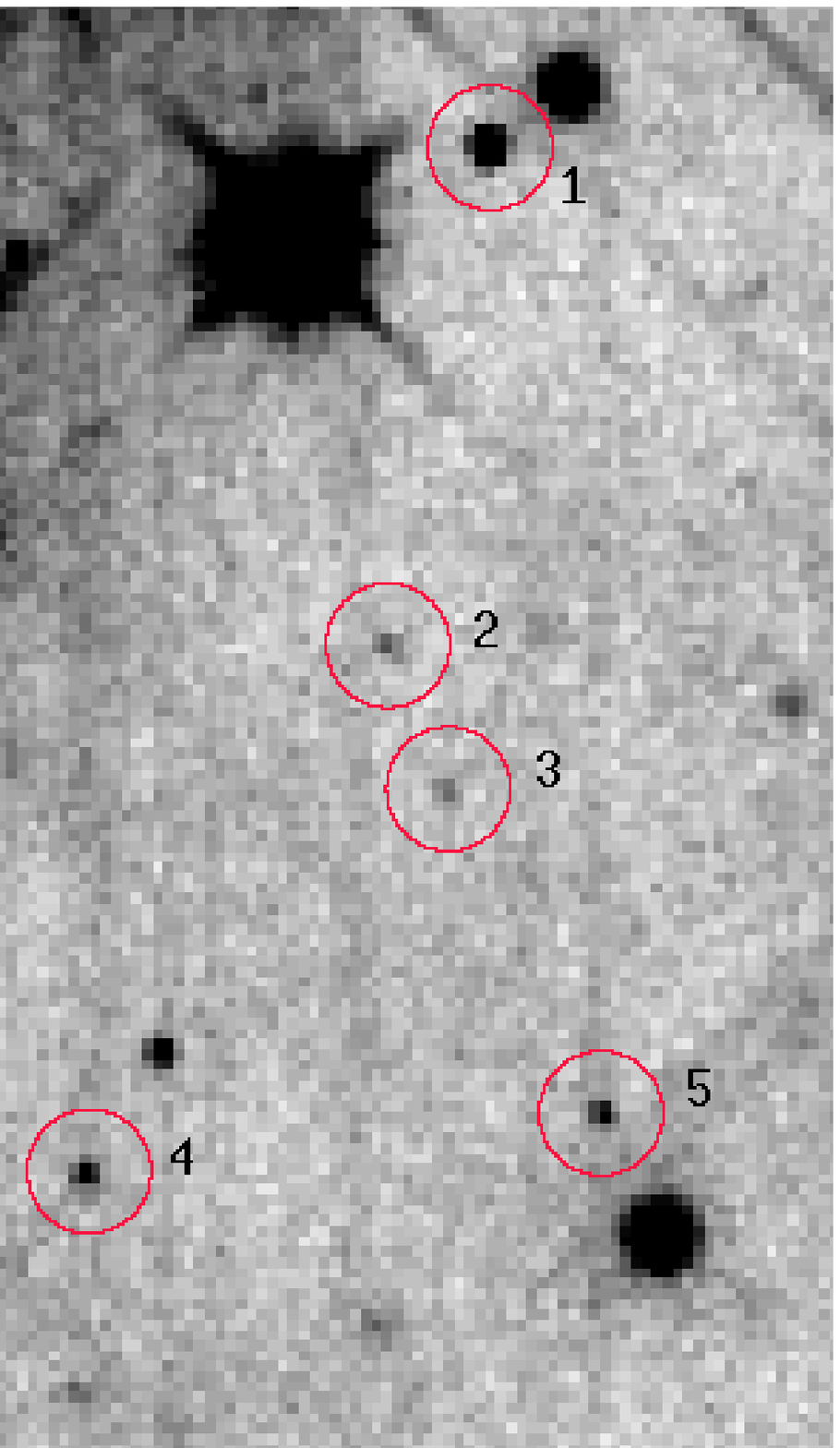}\hspace{0.5cm}
\includegraphics{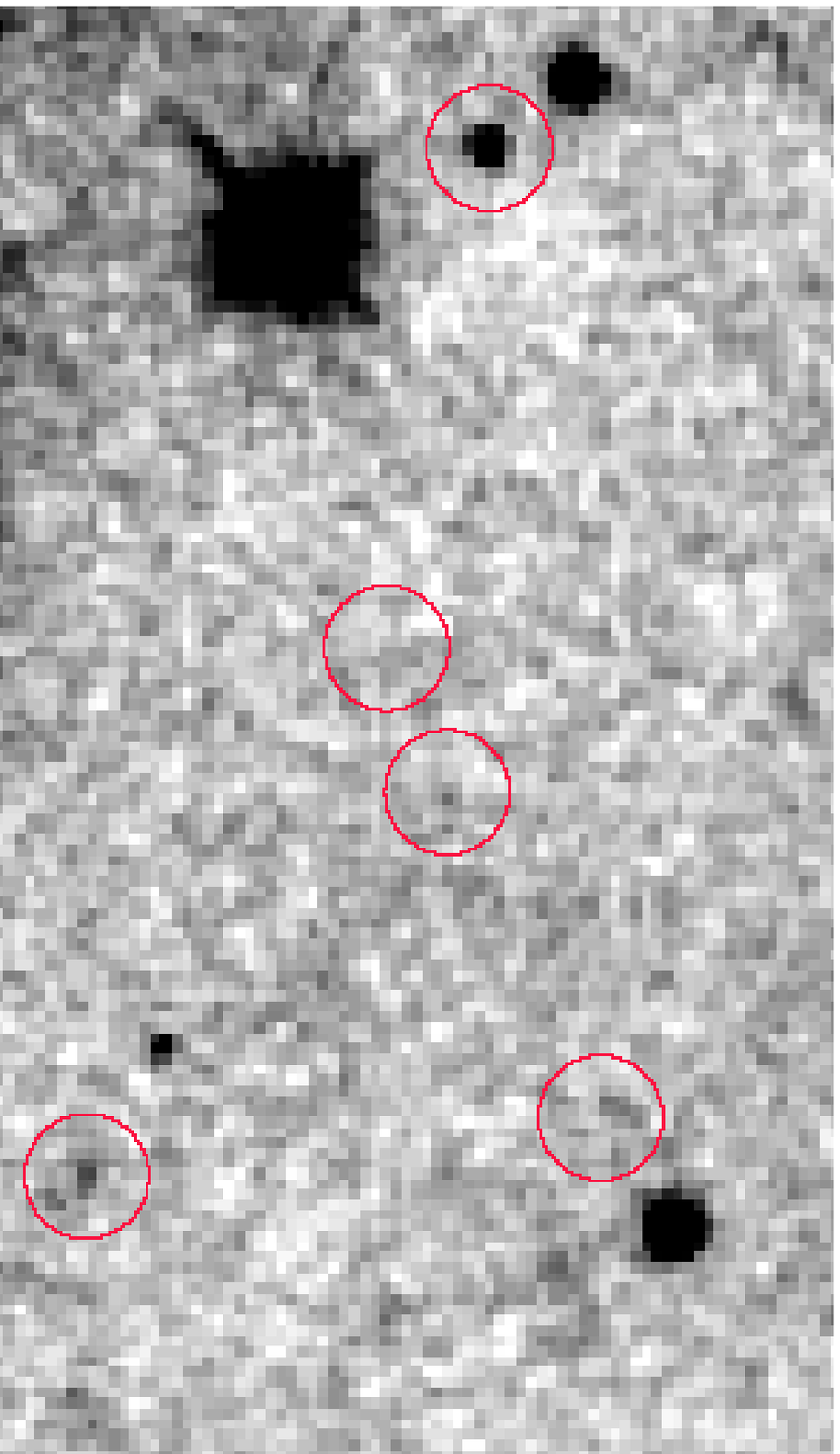}}
\caption{Left panel: section of $23\times 38$ square pixel from the WF4
chip from the second epoch (2001). Right panel: same section, but for
the first epoch (1995). The scale of grays is the same in both images,
which are registered using bona fide MS stars so that no displacement
is expected for cluster members. Whilst it is easy to confirm membership
for for objects nr. 1 and 4 and reject it for nr. 5, nothing can be said
for objects nr. 2 and 3, both at $m_{814}\simeq 27$, since they are
not detected in the first epoch.
}
\label{figure6}
\end{figure*}

If the ``blue hook'' level is defined as the magnitude of the reddest
point in the WD isochrone (in the same way that the MS turn-off is the
bluest), for a cluster age of 8\,Gyr one should expect to find it at
$m_{814}\simeq 26.8$, near the 50\,\% completeness lines, whereas for
10 and 12\,Gyr it should take place at $m_{814}\simeq 27.2$ and
$m_{814}\simeq 27.5$, respectively, i.e. more than half a magnitude
below our $50\,\%$ completeness lines\footnote{Our adopted distance
modulus, $(m-M)_{814}=12.25$, has been used to translate the isochrones
onto the observational plane.}. In any case, because of the large
uncertainties in the data, none of the three isochrones (coloured 
lines in Figure\,5) can be ruled out, nor can their validity be
assessed. Barring the odds that Nature has played tricks on us, i.e.
that she has placed the ``blue hook'' {\em precisely} at the detection
limit, it would be safer to conclude from this simple test that M\,4
must be older than 8\,Gyr. As disappointing as it may seem, this is the
only sound conclusion that one can infer from the available CMD.

\section{The white dwarf luminosity function and the age of the cluster}

With the above caveat in mind, one could turn to the LF of the WDs for more
hints on their age. Several sources of uncertainty come into play here.
Firstly, the LF of the cooling sequence depends on assumptions on the
relationship between the mass of the progenitors and that of the WD,
the time spent by the progenitor on the MS and the initial MF (see e.g.
Chabrier 1999). The WD LF is particularly sensitive to the shape of the
latter since, although the location of the peak of the WD LF depends only
mildly on the slope of the MF (Richer et al. 2000), the rate at which
the LF rises to its peak changes dramatically with it (Chabrier 1999),
thereby casting serious doubts on conclusions based on a WD LF whose
peak is not reached. A sharp maximum is indeed expected in the WD LF of
old stellar systems, a characteristic pile up that marks their age (see
the recent review by Fontaine et al. 2001).  As in the
case of the ``blue hook,'' only when the photometry is deep enough to
clearly reveal the peak of the LF can one safely constrain the
cluster's age.  Otherwise, one can only provide a lower limit to it.

Secondly, the WD LF is only meaningful if it is cleaned of as many
contaminating field stars as possible. The density of objects seen in
the CMD of Figure\,1 between the MS and WD sequence leaves no doubt
that the contamination is strong. To clean the WD sequence one should
then resort to using the same proper motion technique adopted
successfully for MS stars. The problem here is that, the first epoch's
exposures being considerably less deep than those of the second, the
magnitude range over which one can study the WD LF with some
reliability is further reduced. As mentioned above, the dashed lines in
Figures\,1 and 5 mark the magnitude at which, on average, the peak of a
star on the combined images drops below the level corresponding to 5
times the standard deviation of the surrounding background. If the same
criterion were applied to the combined F814W image of the first epoch
(the deepest of the two filters), the same $5\,\sigma$ line should be
drawn at $m_{814}\simeq 25.6$. By relaxing the requirement to just
$3\,\sigma$ for the first epoch alone, one can lower the limit to
$m_{814}\lesssim 26.5$. Below such a threshold, most stars detected in
the second epoch do not have a corresponding match in the first one. We
show this graphically in Figure\,6, where the same area of the combined
F606W+F814W frame is compared at the two epochs, with the same scale of
gray levels. Membership can easily be assessed for object nr. 1 (a MS 
star with $m_{814}\simeq 22$), nr. 4 (a WD with $m_{814}\simeq 25$) and
nr. 5 (a field star with $m_{814}\simeq 25$). However, with magnitudes
around $m_{814}\simeq 27$, the objects marked nr. 3 and 4 are well
visible at the second epoch (left panel) but they are unmeasurable in
the first by any statistically acceptable means at the level of at
least $\sim 2\,\sigma$ in the peak.

\begin{figure}
\resizebox{\hsize}{!}{\includegraphics{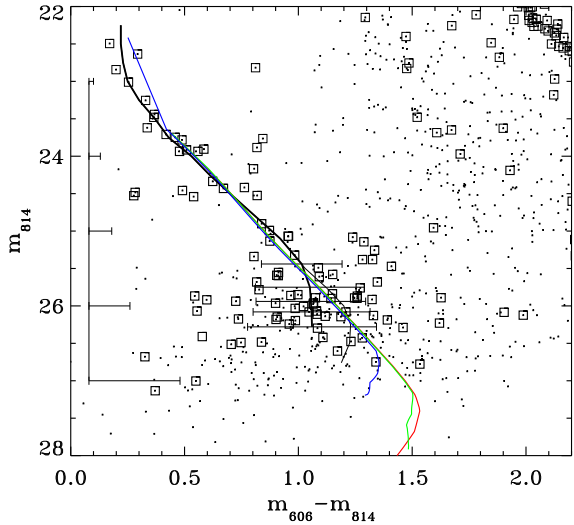}}
\caption{Colour--magnitude diagram of the region occupied by the WDs.
Solid dots indicate all stars from Figure\,5, whereas open squares mark
the objects that moved less than $0\farcs03$ (i.e. $0.3$ WF pixel)
between the two epochs. The thick solid line, showing the WD ridge line
determined as explained in the text, is here compared with the ridge
line of Figure\,5 (thin solid line). Coloured lines as in Figures\,1
and 5}
\label{figure7}
\end{figure}

Furthermore, since the first epoch combined F814W image (despite being
deeper than the F555W one) is the result of only 9 images each of
duration 800\,s, the background is considerably noisier than in the
second epoch F814W frame. It is, thus, possible that one finds there a
lump of counts near the position required to match an object in the
second epoch, yet there is no guarantee that such a lump represent
indeed a true star and, accepting it as such, would require one to make
uncertain assumptions about the physical nature of the sources detected
in the deeper exposure, thus invalidating the basis on which proper
motion studies rest. For this reason, we believe that it is safer to
consider as valid only stars which can be detected as such (at least at
the $3\,\sigma$ level) in the first epoch.

In Figure\,7 we show the CMD of the objects in the WD region. Solid
points mark all the stars detected in the second epoch (the same as
Figure\,5), whereas the squares indicate the objects which were found
in the first epoch as well and whose displacement is smaller than $0.3$
pixel with respect to the nearby bona fide MS stars. (This is the same
selection criterion adopted for Figure\,2). The thick solid line
cutting through the WD sequence is the ridge line, determined as in
Figure\,5. Although the ridge line begins to steepen at $m_{814}
\gtrsim 25.5$, its shape is fully compatible with that of Figure\,5
(thin solid line) within the errors. Consequently, not even in this
case (i.e. by using only the bona fide cluster stars) is it possible to
discriminate amongst the isochrones corresponding to the three ages
mentioned above. The broadening of the cooling sequence with increasing
magnitude, witnessed by the growing size of the error bars, is totally
compatible with the photometric error. Thus, within such an
uncertainty, there is once again no indication of a turn to the blue.

To measure the WD LF, we counted the number of objects in Figure\,7
which are within $4\,\sigma$ of the ridge line, where $\sigma$ is the
photometric uncertainty on the colour indicated on the left-hand side
of the graph. The LF so derived is listed in Table\,1.

\begin{figure}
\resizebox{\hsize}{!}{\includegraphics{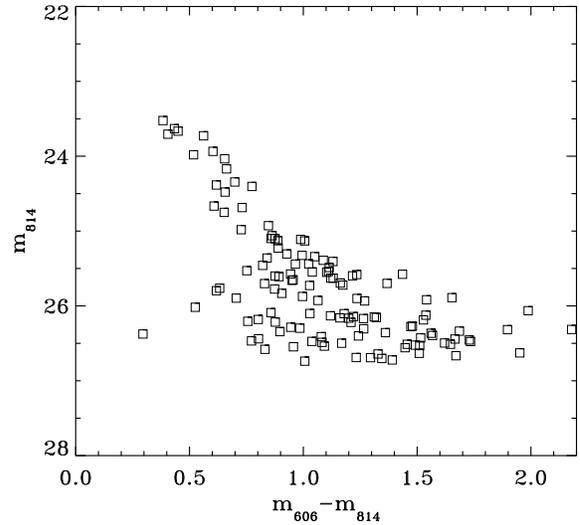}}
\caption{Example of a synthetic CMD, based on the models of PMS02, for
the WDs in a GC of 12 Gyr of age and initial MF index $\alpha=2.3$.
This CMD, which does not include contamination from field stars and is
photometrically complete at all magnitudes, can be compared with that
of Figure\,7}
\label{figure8}
\end{figure}

For an improved comparison between theory and observations, we have
produced a set of synthetic CMDs and LFs for WD stars, as explained in
the following. Firstly, we assume an age and the index of a power-law
MF ($\alpha$) for the cluster. The minimum mass of the WD progenitors
depends on the assumed age, whereas the maximum mass is fixed according
to the detection limit, namely $m_{814}=27$ for the available
photometry. We, therefore, use a Monte Carlo method to generate values
of the progenitor's initial mass, distributed according to the assumed
MF. The relationship between initial and final mass then gives the mass
of the corresponding WD, whose location in the CMD is obtained by
interpolating on the grid of models of PMS02. Finally, photometric
errors are applied according to a gaussian distribution, similar to the
one observed. The extraction of the MS mass values is iterated until
the number of WDs having $m_{814}<26.5$, or $M_{814} < 14.25$ with our
adopted distance modulus, matches the one inferred from the observed
CMD after correction for incompleteness, namely 112. An example of
such a synthetic CMD is shown in Figure\,8, for an age of 12\,Gyr and
$\alpha=2.3$. Synthetic LFs are shown for various values of $\alpha$ in
Figure\,9, and for different ages in Figure\,10. The agreement with the
measured LF (open squares) is excellent, in all cases except 8\,Gyr
(see Figure\,10).

\begin{figure}
\resizebox{\hsize}{!}{\includegraphics{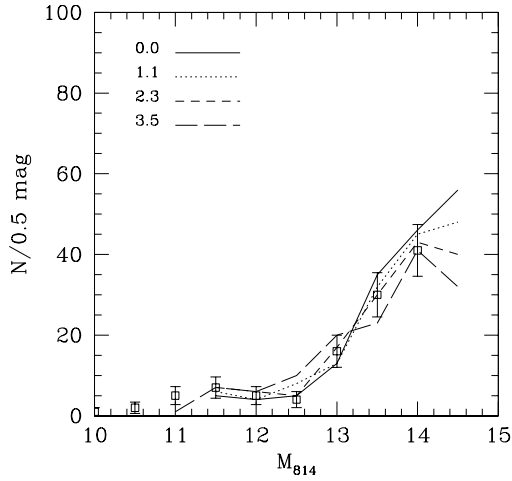}}
\caption{Comparison between the empyrical WD LF of Table\,1 (squares)
and those predicted by our models for 12\,Gyr of age. Each line
corresponds to a different index of the initial MF, as indicated in the
legend. All models fit the data remarkably well}
\label{figure9}
\end{figure}

\begin{table}
\centering
\caption{White dwarf luminosity function. Column (1) gives the
$M_{814}$ magnitude at the centre of each bin, column (2) the number $N_0$ 
of WDs counted in that bin and column (3) the associated photometric
completenes.}
\begin{tabular}{crc}
\hline\hline
\multicolumn{1}{c}{$M_{814}$} &
\multicolumn{1}{c}{$N_0$} &
\multicolumn{1}{c}{compl.} \\
\hline
10.0 &  1 & 1.00 \\
10.5 &  2 & 1.00 \\
11.0 &  5 & 0.95 \\
11.5 &  7 & 0.95 \\
12.0 &  5 & 0.95 \\
12.5 &  4 & 0.95 \\
13.0 & 13 & 0.80 \\
13.5 & 24 & 0.80 \\
14.0 & 27 & 0.65 \\
\hline
\end{tabular}
\vspace{0.5cm}
\label{table1}
\end{table}

\begin{figure}
\resizebox{\hsize}{!}{\includegraphics{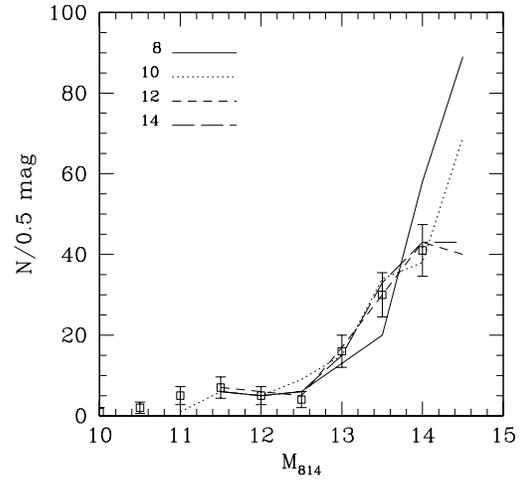}}
\caption{Comparison between the empyrical WD LF of Table\,1 (squares)
and those predicted by our models for an initial MF index $\alpha=2.3$.
Each line marks a different cluster age, as indicated in the legend. The 
models with an age in excess of 8\,Gyr fit the data remarkably well}
\label{figure10}
\end{figure}

The good match between theory and observation provides an important
confirmation of the validity of the theoretical models of the WD
cooling sequence at least up to $M_{814}\simeq 14$. The brightest and
the faintest WD in the synthetic CMD shown in Figure\,8 have a model
mass of $0.5566$ and $0.5901$\,M$_\odot$, respectively, which
correspond to progenitor masses of $0.8660$ and $1.2028$\, M$_\odot$ on
the MS, respectively.  A look at the internal structure of the models
close to the faint end of the observed sequence also provides
interesting information: the Oxygen component of the core is almost
fully crystallised, whilst only 20-25\% (in mass) of the Carbon is in
the solid phase (the rest being liquid). Furthermore, for the faintest
WDs observed (at $M_{814}\simeq 14$), the so called ``convective
coupling'' (see e.g. Fontaine et al. 2001) has just begun. This
phenomenon occurs when the base of the convective envelope reaches the
inner regions of the star, where heat transport is dominated by
electronic conduction. We recall that both the release of latent heat
caused by the liquid/solid phase transition and the convective coupling
induce a substantial decrease of the cooling timescale and, in turn,
affect the age--luminosity relationship. This delay of the cooling is
responsible for the expected pile up of white dwarfs.  Figures\,9 and
10 show that the rate of growth of the observed number of WDs brighter
than $m_{814} \simeq 26.5$ ($M_{814}\simeq 14.25$) does not depend on
the initial MF nor on the age.  Therefore, the very good agreement
between theoretical and observed LF down to the detection limit in M\,4
strongly supports our description of the complex physical phenomena
operating in extreme conditions (high pressure and low temperature).

Unfortunately, in spite of this confirmation, little can be concluded
from this data set about the cluster properties such as its age and
initial MF. In fact, it is immediately evident that, within the
uncertainties, all ages displayed in Figure\,10 are compatible with the
data, although for an age of 8 Gyr a very steep initial MF ($\alpha
>3.5$) would be required to reproduce the WD number counts in the
faintest magnitude bin. Thus, one may tentatively conclude that M\,4 is
older than $9-10$ Gyr (although also these values are model-dependent),
thence placing a slightly improved lower limit with respect to that
already derived by means of the isochrone fitting method. No evidence
of a maximum of the LF is obtained and, in turn, no upper limit for the
age of M\,4 can be set on the basis of these data.

\section{Comparison with previous analyses}

As mentioned in the Introduction, a previous analysis of these same
data by Richer et al. (2002) arrived at the rather different result
that the age of M\,4 is $12.7\pm 0.7$ Gyr (Hansen et al. 2002).
Although we are not able to reconcile completely our result with
theirs, we offer here some considerations that might indicate where the
origin of this discrepancy lies.

The first notable difference is in the assumed distance modulus for
M\,4, which we took from Harris (1996) to be $(m-M)_V=12.83$ with
colour excess $E(B-V)=0.36$, which in turn imply $(m-M)_I=12.25$. The
other widely used catalogue of GC parameters, that of Djorgovski
(1993), gives $(m-M)_V=12.75$ and $A_V=1.24$, which translate into
$(m-M)_I=12.11$. Hansen et al. (2002) used $(m-M)_V=12.51$ based on MS
fitting with subdwarfs of known parallax (see Richer et al. 1997). The
application of such a subdwarf-fitting method requires great caution.
Amongst the many sources of uncertainty, the most severe is related to
the colour shifts that must be applied owing to the differences in
reddening and metallicity between cluster and subdwarf stars (see e.g.
Gratton et al. 1997; Pont et al. 1998; Carretta et al. 2000; Gratton et
al. 2003). For example, an error in the global cluster metallicity
([M/H]) of just $0.1$\,dex implies a change of the estimated distance
modulus of about $0.1$\,mag. In this context, the metallicity adopted
by Richer et al. (1997) in their application of the subdwarf fitting
method, namely $[Fe/H]=-1.3$ without any account for $\alpha$-elements
enhancement, is at odds with the $[M/H]=-1$ value implied by the high
resolution spectroscopy quoted above (Carretta \& Gratton 1997; Carney
1996). By adopting this latter value for the global metallicity of
M\,4, the subdwarf fitting method would give a distance modulus of
$(m-M)_V=12.86$, perfectly in line with the one that we assumed.

The second difference is the adopted photometric system. We simply
calibrated our instrumental magnitudes in the HST in-flight system
(VEGAMAG), which only requires the application of a zero point. Richer
et al. (2002), however, translated their photometry into the
Johnson/Cousin ground-based system and notice a large colour term in
the conversion from F606W to $V$. Thus, it cannot be excluded that,
there not being observed spectra for WD of the type studied here, the
uncertainty in the conversion is very large. On the other hand, we used
for the WDs the models of PMS02, which were generated in the same
VEGAMAG system by using the atmospheres of Bergeron et al. (2001),
whilst Hansen et al. (2002) used the models computed by Hansen (1999)
for the Johnson/Cousin system.

\begin{figure}[t]
\resizebox{\hsize}{!}{\includegraphics{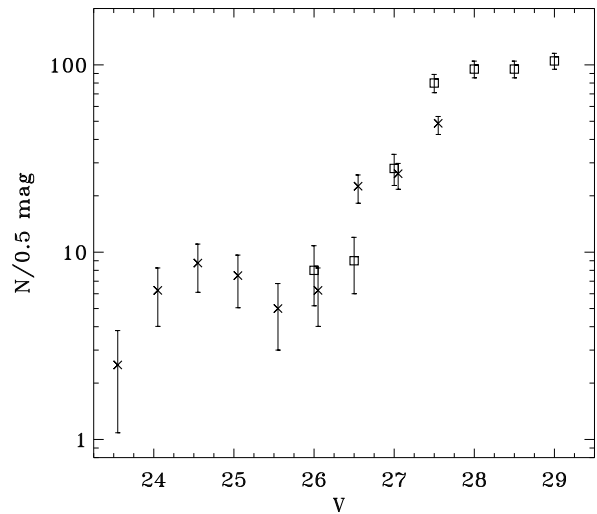}}
\caption{The WD LF as measured by Hansen et al. (2002) in the $V$ band
(squares) is compared here with ours (crosses)}
\label{figure11}
\end{figure}

In order to make it possible to compare our results with those of
Hansen et al. (2002), we measured the WD LF in F606W and attempted to
convert it into Johnson's $V$ band. There not being in these bands
observations of spectro-photometric WDs with the mass, gravity and age
of relevance here, we were forced to base this translation on the
theoretical WD isochrones of PMS02, which, besides mass and gravity,
provide the magnitudes of the cooling WDs in various bands of the
VEGAMAG and Johnson's photometric systems. It is, thus, possible to
derive the $V$ magnitude of each bin of the WD LF directly from the
mass corresponding to the observed F606W magnitude. Given the
remarkably good agreement between the observed and predicted WD cooling
sequence (see Figure\,1) as well as the fact that the F606W and $V$
band span a very similar wavelength range, the colour term is
practically negligible.. The $V$-band WD LF so obtained is shown in
Figure\,11 (crosses), where it can be directly compared with that
published by Hansen et al. (2002, squares).  Both LFs are corrected for
photometric incompleteness (never exceeding 50\,\% for ours). As
expected, since our proper motion selection is more conservative than
that of Hansen et al. (2002; see Figure\,5), our $V$-band LF does not
reach as deep as theirs. The comparison is further hampered by the fact
that Hansen et al. (2002) do not show the LF for $V < 26$. Over the
common magnitude range, however, the two LFs aagree to within $\sim
2\,\sigma$.  We underline here, however, that given the uncertainties
of the conversion from F606W to $V$, Figure\,11 is only meant to show
the comparison between our LF and that of Hansen et al. (2002). The
most reliable LF still remains that in the F814W band.

\begin{figure}[t]
\resizebox{\hsize}{!}{\includegraphics{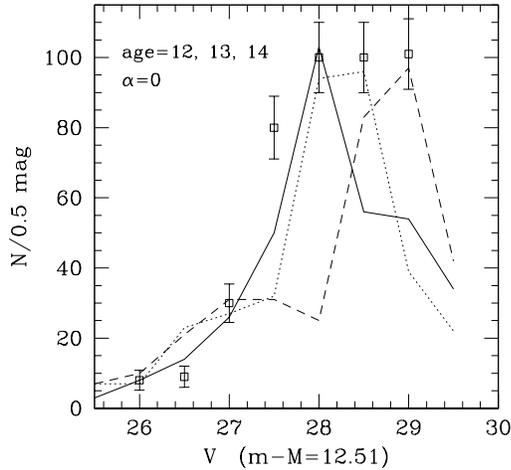}}
\caption{The WD LF as measured by Hansen et al. (2002) in the $V$ band
(squares) is compared here with our model LFs in the same band, which
are evidently unable to reproduce the observations. The least
unsatisfactory fit (dotted line) is obtained for an age of 13\,Gyr but
requires an unphysically flat initial MF ($\alpha=0$)}
\label{figure12}
\end{figure}

Besides directly comparing the two $V$-band LFs, it is also possible to
compare the WD LF as predicted by our models (PMS02) in that band with
the measurements of Hansen et al. (2002). We do so in Figure\,12, where
the squares indicate the WD LF of M\,4 as measured by Hansen et al.
(2002), corrected for incompleteness. We note that the theoretical WD
LF of the same authors reproduces the observations stunningly well for
an age of $12.5$\,Gyr, a distance modulus $(m-M)_V=12.51$ and a
power-law initial MF of intermediate index ($\alpha \simeq 1$). With
the same distance modulus and MF index, the LF predicted by PMS02
departs significantly from the observations, regardless of the adopted
age. In particular, we are unable to reproduce the ratio between the
bright and the faint end of the WD LF as derived by Hansen et al.
(2002). To reduce the discrepancy, we should be forced to use an
unphysically flat MF ($\alpha=0$) which, however, still gives a rather
unsatisfactory fit even for an age of 13\,Gyr (see dotted line in
Figure\,11). However, it did not escape our attention that the
contamination from non cluster members at the faint end of the LF (not
unlikely, given the conditions under which Hansen et al. 2002 have
assessed cluster membership for the faintest objects; see Section\,5)
may indeed mimic the effect of a flat initial MF. In fact, the
photometric completeness of the data-point at $V=29$ does not reach
40\,\%.

A disagreement is, thus, evident between the two sets of theoretical
LFs, which, however, does not correspond to a difference in the cooling
sequences, as witnessed by Figure\,4. To investigate further the origin
of this discrepancy, we have compared the theoretical WD LFs of PMS02
with those tabulated of Richer et al. (2000), which are based on the
models of Hansen (1999). An example is shown in Figure\,13 for an age
of 12\,Gyr, $\alpha=2.3$ and progenitors of solar composition: although
the peaks of the two LFs are displaced by $\sim 0.5$\,mag, the rise and
drop are quite similar. In particular, no plateau is seen at the faint
end such as the one obtained by Hansen et al. (2002). Thus, it seems
that the difference between our fit to the WD LF of M\,4 and that of
Hansen et al. (2002) cannot be attributed to the different theoretical
prescriptions for the physics of the cooling process. We can only
speculate that the discrepancy could be due to differences in the
adopted progenitor ages and/or in the initial--final mass
relationship.

Nevertheless, a general comment on the comparison between theory and
observation can still be done: since the number of WDs is still rising
in the faintest available bin of the observed LF (both in ours in the
F814W band and in that of Hansen et al. in the $V$ band), only a lower
limit to the age can be derived. {\em Once again we emphasise that a
clear identification of the WD LF peak is a mandatory requirement for a
reliable determination of the cluster's age.}

\begin{figure}[t]
\resizebox{\hsize}{!}{\includegraphics{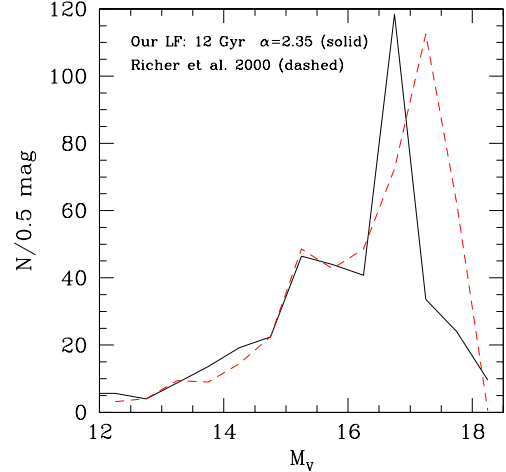}}
\caption{Theoretical WD LFs for cluster age of 12\,Gyr and initial MF
index $\alpha=2.35$ as predicted by our models (PMS02; solid line) and
by those of Richer et al (2000; dashed line). Both models predict a
sharp LF peak, rather different from the plateau measured by Hansen et
al. (2002)}
\label{figure13}
\end{figure}

\begin{figure*}[t]
\resizebox{\textwidth}{!}
            {\includegraphics[bb=18 274 592 648,clip]{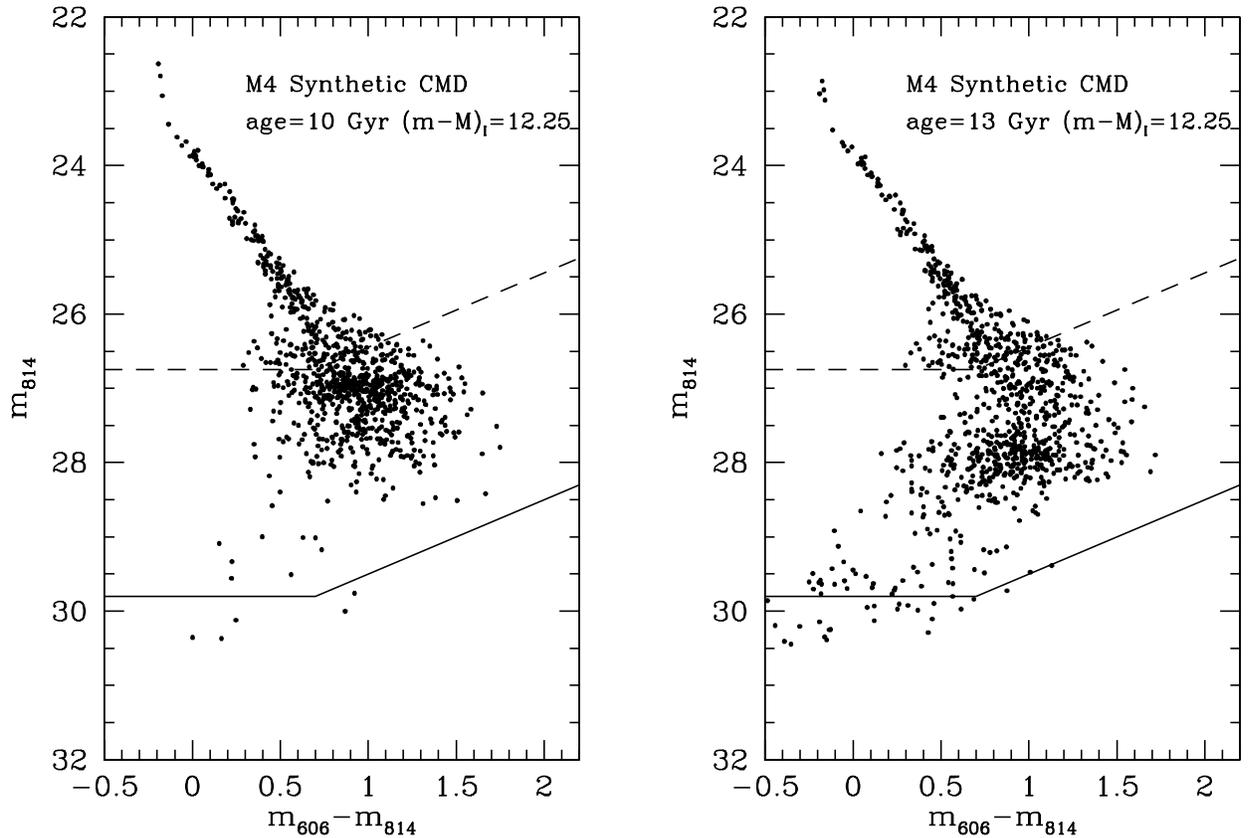}}
\caption{Synthetic CMDs simulating the WD cooling sequence in M\,4, as
it would be observed with the ACS on board the HST, for a cluster age of
10\,Gyr (left-hand panel) and 13\,Gyr (right-hand panel). The ``clump''
of stars responsible for the peak of the WD LF moves to fainter
magnitudes with increasing age}
\label{figure14}
\end{figure*}

Deeper observations, with the sensitivity and accuracy now attainable
with the Advanced Camera for Surveys (ACS) on board the HST, would
dramatically improve this situation. Synthetic CMDs provide us with a
powerful tool to check the effectiveness of measurements at fainter
limits. In Figure\,14 we show a synthetic CMD in which the detection
limits have been moved down to $m_{814}\simeq30$. Photometric errors
have been re-scaled in such a way that they coincide, at the faint
limit, with the photometric errors of the present photometry. As
expected, for the adopted age of 13 Gyr and $\alpha=2.3$, a clump
around $m_{814}=28$ appears. At fainter magnitudes, owing to both the
steepness of the initial MF and the acceleration of the cooling time,
only few stars are predicted, scattered by the large photometric
errors. Since the field of view of the ACS is about $2.3$ times larger
than that of the WFPC2, we have supposed that 260 WDs (instead of 112)
should be found for $m_{814} < 26.5$. Figure\,15 illustrates the
expected variation of the LF for different ages, after proper
normalisation to the total number of stars observed. Interestingly, the
peak of the WD LF is extremely sensitive to the age, shifting at a rate
of $\sim 0.5$ mag/Gyr (for ages ranging between 10 and 13\,Gyr),
whereas the corresponding variation of the MS turn-off luminosity is
only $\sim 0.1$ mag/Gyr (for an age of 12\,Gyr). This implies that the
age estimated by means of the WD LF is considerably less sensitive to
the uncertainty on the distance than the latter. On the basis of these
simulations, we conclude that the peak of the WD LF in M\,4 should be
clearly identifiable by means of the ACS in a reasonable exposure time,
if the cluster's age were $< 13.5$\,Gyr as expected from the recent
WMAP measurements.

\begin{figure*}
\resizebox{\textwidth}{!}
            {\includegraphics[bb=18 274 592 600,clip]{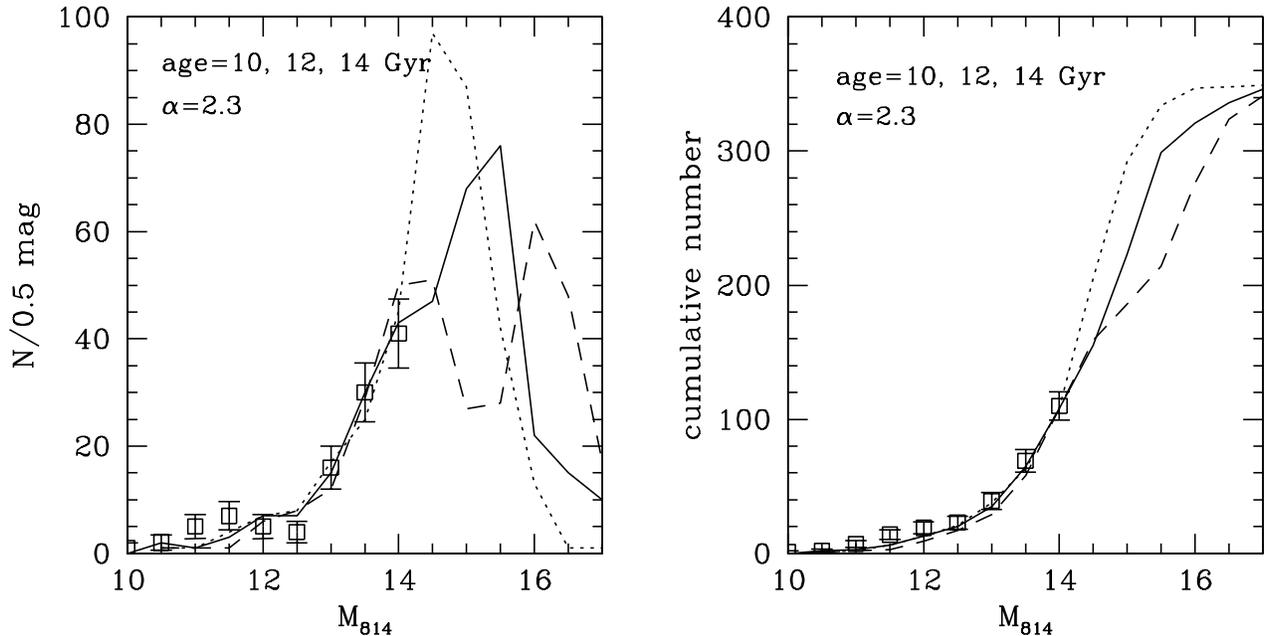}}
\caption{The sensitivity of the ACS on board the HST makes it possible
to access the magnitude range above the limit currently set by the
WFPC2 observations (squares) at $M_{814}\simeq 14$, where the WD LF is
most sensitive to the cluster's age. Theoretical differential and
cumulative WD LFs are shown, respectively in the left- and right-hand
side panel, for an initial MF index $\alpha=2.3$ and the ages as
indicated in the legend}
\label{figure15}
\end{figure*}

\section{Summary}

The main results of this paper can be summarised as follows.

\begin{enumerate}

\item
We have reduced and analysed a set of deep observations of the GC M\,4
obtained in 2001 with the WFPC2 camera on board the HST. This data set
includes 98 images in the F606W filter and 148 images in the F814W
band, each of duration 1300\,s. All the frames have been subjected to
the standard HST pipeline which resulted in two calibrated, registered
and coadded images (one per filter) on which standard aperture
photometry was run. We show that stars can be reliably measured down to
the 50\,\% completeness limit at $m_{606}\simeq 28$, $m_{814}\simeq
27$. Above magnitudes brighter than $m_{814}\simeq 19$ the photometry
is not reliable because of saturation.

\item
We derive a CMD (Figure\,1) which reveals a narrow and well defined
stellar MS extending to $m_{814}\simeq 23$, where it becomes
indistinguishable from field stars. The WD cooling sequence is also
visible and extends from $m_{814}\simeq 22.5$ through to the detection
limit where it broadens considerably due to the increasing photometric
uncertainty and where field star contamination is most severe. The
signature of the Galactic disc and bulge is clearly visible in the CMD
as a cloud of points occupying the region between the WD cooling
sequence and the MS.

\item
The shape of the cluster MS differs from that predicted by the models of
Baraffe et al. (1997) for any choice of distance modulus. This shortcoming
in the theoretical description of low-mass stars has already been noted
by Bedin et al. (2001) and stems from the inadequacy of the available
treatment of the TiO molecule opacity (Chabrier 2001) in the F606W and
bluer bands. We derive the LF of MS stars in the F814W band, where
uncertainties in model atmospheres are smaller, by imposing a colour
selection around the MS ridge line on the CMD down to $m_{814}=23.5$ or
$M_{814}=11$ (Figure\,2). We show that the LF is not compatible with a
single power-law underlying MF, regardless of the adopted index. Instead,
the observations are well reproduced by a tapered power-law MF with peak
mass $m_p=0.35$\,\Msolar, index $\alpha=2.1$ and tapering exponent
$\beta=2.7$ (Figure\,3). For comparison, the average parameters for 12
halo GCs studied with the HST are $m_p=0.34$, $\alpha=2.3$, $\beta=2.6$.

\item
Adopting the canonical distance modulus for M\,4 of $(m-M)_I=12.25$, our
theoretical WD isochrones overlap remarkably well with the observed
cooling sequence (Figure\,5). The data, however, do not reach the
magnitude domain where isochrones of different ages depart significantly
from one another, nor that where the rapid turn to the blue of the
sequences (``blue hook'') caused by collision-induced absorption is
expected to take place. Therefore, only a lower limit of $\gtrsim 8$\,Gyr
can be set to the age of M\,4 from the CMD alone.

\item
To set more stringent constraints on cluster membership than those
allowed by the colour of the stars in the CMD and, thus, derive accurate
LFs for both MS stars and WDs, we have reduced and analysed a set of
shallower observations of the same cluster field obtained in 1995 with
the WFPC2 with the aim of measuring proper motions. The considerably
shorter exposure times (particularly in the F814W band) forced us to
restrict this study to objects brighter than $m_{814}\simeq 26.5$, since
fainter objects are not detectable by any statistical means at the level
of at least $3\,\sigma$ and, therefore, their presence, position and
nature cannot be securely confirmed (Figure\,6). Above this limit,
cluster members can rather easily be separated from field stars since
their displacement amounts to $\sim 0\farcs1$ between the two epochs,
or a full WF pixel.

\item
The LF of MS stars, selected this time via their proper motions, agrees
remarkably well with that obtained through colour selection in the CMD
over the common magnitude range. The former, however, extends further to
$m_{814}= 25$ or $M_{814}=12.25$ where the number of MS star is
consistent with being zero. The underlying MF that best fits this LF is
the same tapered power-law distribution mentioned above, thus
suggesting that, although important for obtaining information on the
faint MS end, the availability of proper motions does not substantially
improve our understanding of the number and distribution of low mass
stars. In fact, we stress that the known shortcomings in modelling the
atmosphere of very low mass objects (see Figure\,1) make it difficult
to decide whether and where one has reached the bottom of the MS.

\item
On the other hand, knowing which objects belong to M\,4 makes it
possible to study the LF of the WDs, whose location in the CMD is
severely contaminated by field stars. However, the shallower photometric
depth of the first epoch's data limits this investigation to $M_{814} <
14.5$, thereby preventing us from exploring the domain where the WD LF is
most sensitive to the cluster's age and to its initial MF. Our models
show that MF indices in the range $0< \alpha < 3.5$ and ages in the
range $8 - 14$\,Gyr are all consistent with the data (Figures\,9 and 10),
although for an age of 8\,Gyr an unrealistically steep MF ($\alpha >
3.5$) would be required to reproduce the number counts in the faintest 
bin of the LF. On the basis of these data and our models, we can set a 
lower limit of $\gtrsim 9$\,Gyr to the age of M\,4. We underline here 
that an upper limit to the age can only be set when the sharp maximum of 
the WD LF is detected which results from the characteristic pile up of 
old WDs along their cooling sequence (Fontaine et al. 2001). No such 
feature is seen in the presently available data and, therefore, only a 
lower limit to the age can be set.

\item
We have compared our results with those of Hansen et al. (2002) who,
using the same data set, derived for M\,4 an age of $12.7 \pm 0.7$\,Gyr
from the WD LF in the $V$ band. We are unable to reproduce
satisfactorily their observations with our theoretical $V$-band LF for
any physically meaningful value of the WD progenitors' initial MF
index $\alpha$. In particular, we are not aware of models predicting a
plateau in the LF such as the one obtained by Hansen et al. (2002) for
$V\gtrsim 28$. The latter could be the consequence of an excess of
spurious WDs at the faint end of the LF arising from uncertainties in
the proper motions of objects which are difficult to detecte in the
first epoch. Nevertheless, since the WD LF of Hansen et al. (2002) is
still rising at the faintest bin, no upper limit to the age can anyhow
be set.
 
\item
We have used synthetic CMDs to simulate deeper GC observations such as
those today attainable with the ACS on board the HST. The combination
of a wide field of view and excellent sensitivity in the F606W and
F814W bands make it possible for this instrument to study the physical
properties of old WDs, to detect the ``blue hook'' in their cooling
sequences and to locate with accuracy the peak of their LF, thereby
determining their age, if M\,4 is younger than $\sim 13.5$\,Gyr as the
recent WMAP measurements indicate. Since the peak of the LF moves by
$0.5$\,mag per Gyr of age (for ages ranging between 10 and 13\,Gyr;
Figure\,15), the latter can be measured with an accuracy comparable
with that of $H_0$, thus setting robust cosmological constraints to the
time of GC formation.

\end{enumerate}

\begin{acknowledgements}
In this work we have made extensive use of the ESO/ST-ECF archival facilities.
It is our pleasure to thank an anonymous referee for providing us with
comments that have considerably strengthened the presentation of this
work.
\end{acknowledgements}

\end{document}